\documentclass[conference]{IEEEtran}

\usepackage[letterpaper, top=54pt, bottom=54pt, left=54pt, right=54pt]{geometry}

\IEEEoverridecommandlockouts                              

\usepackage{multirow}
\usepackage{caption}
\usepackage{cite}
\usepackage{amsmath,amssymb,amsfonts}
\usepackage{algorithmic}
\usepackage{algorithm}
\usepackage{graphicx}
\usepackage{textcomp}
\usepackage{xcolor}
\usepackage{adjustbox}
\usepackage[hidelinks]{hyperref}

\def\BibTeX{{\rm B\kern-.05em{\sc i\kern-.025em b}\kern-.08em
    T\kern-.1667em\lower.7ex\hbox{E}\kern-.125emX}}

\usepackage{cuted}
\usepackage{subcaption}
\usepackage{array}
\usepackage{comment}
\usepackage{balance}

\usepackage{array}
\usepackage{booktabs}
\usepackage{multirow}
\usepackage{caption}

\newcolumntype{C}[1]{>{\centering\arraybackslash}m{#1}}

\title{\LARGE \bf
Analysis of Motor Signatures of Social Adaptation\\in Autism for Efficient Human-Centric Systems
}

\author{\IEEEauthorblockN{
Lara Pereira\IEEEauthorrefmark{1},
Teresa Sousa\IEEEauthorrefmark{2},
Miguel Castelo-Branco\IEEEauthorrefmark{2} and
João Ruivo Paulo\IEEEauthorrefmark{1}}

\IEEEauthorblockA{\IEEEauthorrefmark{1}
Institute of Systems and Robotics, University of Coimbra, Portugal\\
Emails: \{lara.pereira, jpaulo\}@isr.uc.pt}

\IEEEauthorblockA{\IEEEauthorrefmark{2}
Coimbra Institute for Biomedical Imaging and Translational Research (CIBIT)\\
of the University of Coimbra, Portugal\\
Intelligent Systems Associate Laboratory (LASI), Portugal\\
Institute of Physiology, Faculty of Medicine, University of Coimbra, Portugal\\
Emails: tsousa@uc.pt; mcbranco@fmed.uc.pt}

\thanks{© 2026 IEEE. Personal use of this material is permitted. Permission from IEEE must be obtained for all other uses, in any current or future media, including reprinting/republishing this material for advertising or promotional purposes, creating new collective works, for resale or redistribution to servers or lists, or reuse of any copyrighted component of this work in other works.}  
}

\begin{document}

\maketitle
\thispagestyle{empty}
\pagestyle{empty}

\begin{abstract}
Dance imitation integrates motor planning, sensorimotor integration, and social cognition, offering a sensitive framework to characterize motor behavior in autism. In this work, we explore a computational analysis framework to identify potential biomarkers that allow the design and development of improved medical and human-machine systems. We analyzed 3D motion capture data from autistic and neurotypical adults performing dance imitation under solo and socially-framed duo conditions. Methodologically, using Dynamic Time Warping, we quantified movement consistency and propose the Social Context Sensitivity Index (SCSI) to measure modulation of variability by social framing. These features were then used on a classifier to discriminate subjects into autistic or neurotypical groups. Results show that neurotypical adults exhibited increased movement variability in socially-framed imitation, especially in upper and lower limbs, whereas autistic adults maintained consistent movement across contexts. Classification achieved 79.2\% balanced accuracy in distinguishing groups. These findings suggest that social context sensitivity in motor imitation constitutes a robust biomarker of autism-related motor behavior, highlighting the importance of social modulation in motor assessments and informing the development of inclusive human-centric technologies.
\end{abstract}

\begin{IEEEkeywords} autism; dance imitation; social context sensitivity; motion analysis. 
\end{IEEEkeywords}

\section{INTRODUCTION}
Autism is a neurodevelopmental condition marked by a combination of behavioral and cognitive impairments, such as differences in social communication, sensory processing, and repetitive and restrictive behaviors~\cite{Lordan2021}. 
Current diagnostic practice relies on structured clinical behavioral observation and caregiver-reported instruments, a process that is subjective and time-intensive~\cite{Vabalas2020}. 
The absence of objective, quantifiable biomarkers contributes to diagnostic delays, also affecting the effectiveness of interventions \cite{Vabalas2020, Simeoli2024}. Developing computational tools capable of characterizing autism-related differences from measurable behavioral signals represents an important step toward more accessible and consistent assessment.

Motor differences are among the most consistently reported features of autism, present throughout life and spanning a range of behaviors including gait, postural control, manual dexterity, and imitation~\cite{Fournier2010, Dowell2009, daSilva2025}. 
Rather than being isolated biomechanical artifacts, motor systems are deeply intertwined with social cognition, action observation, and sensorimotor prediction, all of which are central to the autism phenotype~\cite{Gowen2013,Duarte2022}. Despite this, motor behavior has received comparatively little attention as a source of objective biomarkers in automated assessment pipelines.

Imitation tasks are particularly well-suited for studying the intersection of motor and social processing in autism. Imitation requires the integration of observed biological motion with one's own motor system, engaging action observation networks and predictive motor planning simultaneously~\cite{McEllin2018,Latrche2024}. 
Crucially, the social framing of the observed movement may modulate the motor response differently in autistic and neurotypical individuals, reflecting differences in social motivation and biological motion processing~\cite{Nackaerts2012}.

Dance imitation offers a naturalistic, full-body paradigm that engages rhythmic coordination, whole-body postural control, and expressive movement, dimensions that are rarely examined together in autism motor research. 
In more detail, dance can be presented in both solo and socially-framed variants without changing the fundamental motor demands of the task, enabling a within-paradigm comparison of how social context modulates motor behavior. This design allows group differences attributable to social processing to be dissociated from those attributable to general motor ability, which is a key methodological advantage over single-condition paradigms. 

In this work, we propose a computational analysis framework that identifies potential autistic biomarkers for the design and development of medical and human-machine systems. The analysis uses 3D motion capture data from autistic and neurotypical adults performing dance imitation tasks under three conditions: a baseline free movement condition, a solo dance imitation condition, and a socially-framed duo dance imitation condition. 
Movement consistency across trials is quantified using Dynamic Time Warping (DTW), and a novel per-participant measure - the Social Context Sensitivity Index (SCSI) - is proposed to capture how social framing of the imitation stimulus modulates movement consistency. 
The three-condition progressive design allows us to disentangle general motor differences from those specific to imitation, and further from those specific to socially-framed imitation.

The main contributions of this work are:
\begin{enumerate}
    \item A progressive analysis framework that separates baseline motor behavior, imitation-specific differences, and social context modulation in dance imitation.
    \item The Social Context Sensitivity Index (SCSI), a novel per-participant biomarker quantifying how social framing of the imitation stimulus modulates movement consistency across joint groups.
    \item A classification pipeline combining DTW consistency and SCSI under Leave-One-Subject-Out cross-validation with nested hyperparameter tuning, achieving 79.2\% balanced accuracy and demonstrating that the contrast between solo and socially-framed imitation is the key discriminative signal.
\end{enumerate}

%

\section{RELATED WORK}
\label{sec:relatedwork}

While traditional kinematic assessments for autism have focused primarily on gross motor differences, such as postural instability and gait anomalies~\cite{McCleery2013, daSilva2025}, recent state-of-the-art models for skeletal sequence analysis are increasingly dominated by deep learning architectures, including spatio-temporal graph convolutional networks~\cite{Yan2018}. 
While these models excel at high-capacity pattern recognition, their ``black-box'' nature fundamentally limits clinical interpretability, making it difficult to isolate the exact biomechanical factors driving a prediction~\cite{Taye2023}. 
As an alternative approach, classical machine learning pipelines driven by engineered kinematic features have been proposed \cite{Yan2018,laraieee,datasetarticle}. However, the underlying socio-motor dynamics are not considered in previous works, failing to capture how an individual's movement changes in response to social stimuli.

To build a system capable of measuring this social modulation, we first require a highly interpretable baseline metric. Therefore, to prioritize clinical transparency, our pipeline relies on similarity measures, such as DTW~\cite{Liu2024}. This can compute the direct non-linear divergence between kinematic trajectories, accommodating natural variations in movement speed and onset timing. Within health informatics, DTW has been robustly applied to quantify movement consistency in conditions such as Parkinson's disease and cerebral palsy~\cite{Barth2013, Lee2024}. In our pipeline, DTW is distinctly used not to compare a participant against a reference template, but to compute a within-participant measure of trial-to-trial motor consistency, a highly interpretable clinical metric.

Finally, a critical systemic limitation of existing motor analysis pipelines is their reliance on isolated, non-social actions~\cite{Bolis2018}. While movements in neutral environments are typically optimized for task completion, energy cost, and fine motor control, social contexts introduce a flexible, top-down modulation that systematically alters movement kinematics~\cite{Trujillo2018}. Observing socially-embedded biological motion naturally increases functional movement variability as the observer's sensorimotor system adapts to the stimulus~\cite{Iacoboni2009, Cattaneo2007}. 
Within autism research, this context-dependency reveals a consistent behavioral dissociation: while autistic individuals typically reproduce overarching action goals successfully, they often exhibit reduced fidelity in copying specific movement styles and kinematics~\cite{Forbes2016}. Crucially, this kinematic variance is heavily influenced by external stimuli. Emotional and social contexts fundamentally shape both the perception and production of biological motion, higher-order information where autistic individuals frequently demonstrate atypical processing~\cite{Todorova2019}. 
Existing analytical frameworks are largely unable to capture these nuanced kinematic distinctions because they do not model the dynamic interaction between the participant and the social content of the stimulus~\cite{co}.

Our work addresses this critical system-level gap. By introducing the Social Context Sensitivity Index (SCSI) as a cross-condition kinematic contrast built upon our DTW pipeline, we propose an analytical metric that explicitly quantifies differential motor consistency when a participant transitions from a neutral to a socially-embedded imitation task.

\section{METHODS}
\label{sec:methods}

\begin{figure}[!t]
\centering
\includegraphics[width=0.999\columnwidth]{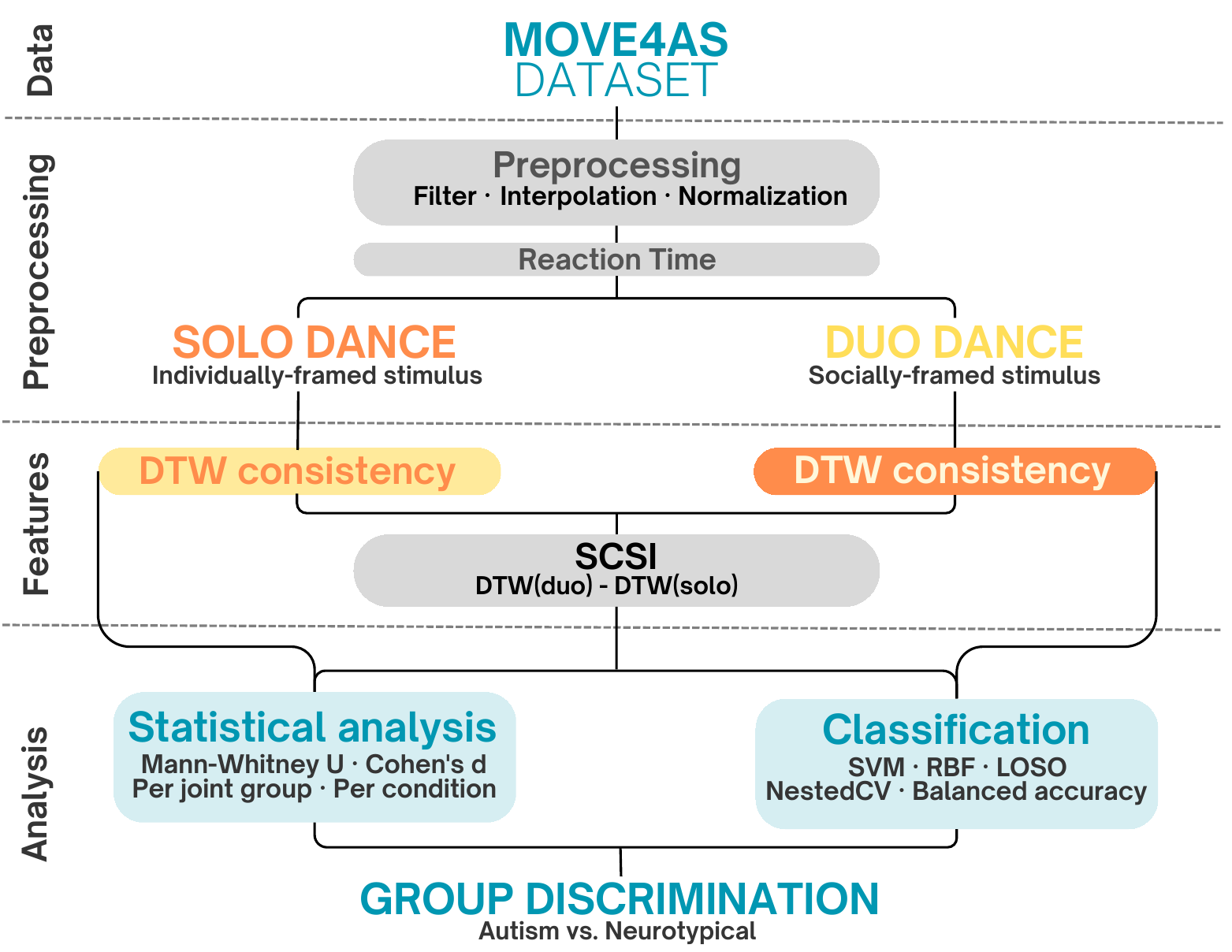}
\caption{Proposed analysis pipeline. Motion capture data are preprocessed and split by condition. DTW consistency is computed per condition and combined into the SCSI. All features feed into statistical analysis and SVM classification.}
\label{fig:pipeline}
\end{figure}

To capture these contextual dynamics from full-body skeletal sequences, we designed the comprehensive computational pipeline illustrated in Fig.~\ref{fig:pipeline}. The architecture is divided into four primary stages: Data, Preprocessing, Features, and Analysis. The pipeline begins with the preparation of the Move4AS dataset, which is partitioned into two distinct contextual streams: a solo dance task featuring an individually-framed stimulus, and a duo dance task featuring a socially-framed stimulus. Following standard preprocessing (filtering, interpolation, and normalization) and the establishment of a reaction time baseline, within-condition DTW consistency is calculated for both tasks. These independent metrics are then synthesized into the SCSI. Finally, the analysis module employs both statistical testing and machine learning classification to achieve robust group discrimination between autistic and neurotypical adults.

\subsection{Move4AS Dataset}
This work leverages the Move4AS~\cite{dstdoi, datasetarticle} multimodal dataset, collected by our team group at the Institute of Systems and Robotics in collaboration with the Portuguese Association for Developmental Disorders and Autism (APPDA). Designed to allow the study of action-perception and social-emotional dynamics. It comprises simultaneous electroencephalography (EEG) and motion capture data.
The present work focuses on the 3D motion data derived from the dance imitation tasks, leveraging their naturalistic, full-body paradigm and embedded social context manipulation.

\subsubsection{Participants}
This dataset comprises a control group of 20 neurotypical individuals, including 13 males and 7 females with a mean age of $25.9 \pm 3.8$ years, alongside a clinical group of 14 autistic individuals, comprising 12 males and 2 females with a mean age of $27.0 \pm 6.8$ years. 

\subsubsection{Data Collection}
Motion capture data were recorded using an OptiTrack Flex 3 system configured with 10 cameras positioned around a $5 \times 5$\,m capture area. Participant movements were tracked using 37 reflective markers, enabling automated reconstruction of a full-body skeletal model. Visual instructions were presented on a 49-inch LCD display, which also provided temporal cues to guide participant performance. Data acquisition was managed through a dedicated workstation running OptiTrack Motive software. Temporal synchronization between motion data and experimental events was achieved by integrating trigger signals into the recorded data stream.

\paragraph{Experimental Paradigm}
The participants performed three dance movement conditions:

\begin{itemize}
    \item \textbf{Solo Dance Imitation:} Participants imitated an individually-framed dance sequence presented as point-light animation.
    \item \textbf{Body Shake:} Participants performed free rhythmic movements following a text instruction, without imitation.
    \item \textbf{Duo Dance Imitation:} Participants imitated a socially-framed dance sequence presented as point-light animation.
\end{itemize}

Each condition consisted of 10 trials per block, in random order. Participants performed the tasks in four recording blocks for the control group and three recording blocks for the autistic group.
Each trial followed a structured timeline: a fixation cross (1\,s), followed by the instruction period (3\,s), an auditory beep signaling movement onset, 4\,s of movement execution, and a second beep followed by a return phase. Motion capture data were recorded continuously throughout.

\subsection{Data Preprocessing}

Raw motion capture data were preprocessed using the pipeline established in the Move4AS dataset descriptor~\cite{datasetarticle}. Following visual screening to exclude trials with compromised tracking, the sequences were low-pass filtered (4th-order Butterworth) to eliminate high-frequency noise. Each trial was segmented into a 7-second window (3\,s instruction, 4\,s execution). To correct for hardware-induced variable sampling rates, trials were temporally aligned via linear interpolation, yielding a uniform 490-frame sequence per trial.

To ensure that the subsequent analysis isolated genuine motor behavior, rather than anatomical disparities, a segment-wise skeletal standardization was applied, scaling individual bone lengths to a global average~\cite{normalizationscale_Zanfir2013}. Finally, position and orientation normalizations were performed~\cite{normalization_pos_orient_Sedmidubsky2017}. The root joint's (hip) coordinates were used as the global reference's origin $(0,0,0)$, and a planar rotation around the vertical $z$-axis was applied to align the initial transverse pelvic vector strictly with the positive $x$-axis (facing the $y$-direction). This ensured all trajectories shared a common spatial reference, isolating pure kinematic execution from baseline spatial offsets.

\subsection{Reaction Time}
Previous research indicates that the perceptual-motor phase is altered in autistic individuals when processing biological motion \cite{Nackaerts2012}. Therefore, to isolate this cognitive phase of motor planning from active kinematic execution, movement onset timing was characterized by computing the reaction time for each trial. This was defined as the latency between the auditory cue and the first frame at which sustained movement was detected.
To isolate limb and trunk movements from global body translation, the 3D Cartesian coordinates of all 21 joints were first re-centered relative to the pelvis at each frame. 

Velocity was computed as the magnitude of frame-to-frame spatial displacement for each joint. The maximum velocity observed across all joints at a given frame, denoted as $V_{\text{max}}(t)$, served as the global onset signal. 

To account for individual resting tremor or minor postural sway, and to avoid the temporal biases associated with absolute kinematic thresholds \cite{Brenner2019}, a participant-specific movement threshold, $\tau$, was established:

\begin{equation}
    \tau = \mu_{\text{base}} + 3\sigma_{\text{base}}
\end{equation}

\noindent where $\mu_{\text{base}}$ and $\sigma_{\text{base}}$ represent the mean and standard deviation of $V_{\text{max}}(t)$ during the instruction period, established as the baseline. 

Reaction time was identified as the first frame, after the beep ($t_{\text{beep}}$), where the maximum joint velocity exceeded this threshold for two consecutive frames. This dual-frame requirement was to filter out transient noise and false detections. Formally, the onset frame $t_{\text{onset}}$ was defined as:

\begin{equation}
    t_{\text{onset}} = \min \{ t > t_{\text{beep}} \mid V_{\text{max}}(t) > \tau \land V_{\text{max}}(t+1) > \tau \}
\end{equation}

The final reaction time (in milliseconds) was computed as the latency between $t_{\text{beep}}$ and $t_{\text{onset}}$ divided by the frame rate. 
The reaction times were then averaged across all valid trials per participant. 
Given the unequal sample sizes and variances between groups, Welch's independent samples $t$-test was employed for statistical comparisons.

\subsection{Feature Extraction}
Motor features were extracted from motion data for each participant in the solo and duo conditions. To avoid influence from variable reaction times at movement onset, features were computed in the last 3 seconds of the execution period.

\subsubsection{DTW Movement Consistency}
Intra-participant movement consistency was quantified using Dynamic Time Warping (DTW) with a Euclidean distance metric. Because human motor execution inherently contains temporal jitter (e.g., natural variations in movement speed across trials), standard linear frame-by-frame comparisons often mischaracterize true spatial variance. DTW mitigates this by non-linearly warping the time axes to optimally align the sequences, effectively isolating true kinematic variability from simple phase shifts~\cite{Switonski2019}. 

For each participant, pairwise DTW distances were computed across all valid trials using the raw 3D joint coordinates. The participant's mean DTW distance across these pairwise comparisons served as their overall consistency score. 
This approach has been recently demonstrated to be an effective measure of intra-individual ``motor noise'' in autistic populations~\cite{Mandelli2024}, where lower DTW values indicate a highly consistent, stereotyped repetition of the motor plan.

For a participant with $N$ valid trials, the intra-participant consistency score, $C$, was calculated as the mean of all unique pairwise DTW distances:
\begin{equation}
    C = \frac{2}{N(N-1)} \sum_{i=1}^{N-1} \sum_{j=i+1}^{N} \text{DTW}(T_i, T_j)
\end{equation}
where $T_i$ and $T_j$ represent the multidimensional kinematic time series of trials $i$ and $j$, respectively. 

To evaluate group-level kinematic consistency, individual scores ($C$) were aggregated to compute the overall mean consistency for both the clinical and control cohorts. 

For the classification models, this DTW consistency was computed separately for three joint groups: upper body (shoulders, arms, hands), core body (hips, abdomen, chest, neck, head), and lower body (thighs, shins, feet).

\subsubsection{Social Context Sensitivity Index}
To capture how participants adapted to the presence of a partner, a Social Context Sensitivity Index (SCSI) was defined as the difference in DTW consistency between the socially-framed duo condition and the solo imitation condition:

\begin{equation}
    \text{SCSI} = \text{DTW}_{\text{duo}} - \text{DTW}_{\text{solo}}
    \label{eq:scsi}
\end{equation}

A positive SCSI indicates greater movement variability in the socially-framed condition relative to solo imitation.

\subsection{Statistical Analysis}

Group differences in movement consistency between the clinical and control groups were statistically evaluated. To account for potential non-normal distributions in the 3D motion data, group comparisons were assessed using the non-parametric Mann-Whitney U test (two-tailed). Independent samples t-tests were also computed to confirm the robustness of the findings. 

Effect sizes were quantified using Cohen's $d$ to determine the magnitude of the group differences (interpreted as small $d < 0.5$, medium $d < 0.8$, and large $d \ge 0.8$). Furthermore, 95\% confidence intervals (CIs) for the group means were generated using a bootstrapping procedure with 1,000 resamples. The threshold for statistical significance was set at $\alpha = 0.05$. 


\subsection{Classification}

A binary classification framework was implemented to evaluate the discriminative power of the extracted motor features. Classification was performed using a Support Vector Machine (SVM) with a radial basis function (RBF) kernel and balanced class weights to account for the
unequal group sizes ($n_{\text{clinical}} = 12$, $n_{\text{control}} = 18$).
Features were standardized to zero mean and unit variance within each training fold to prevent data leakage.

Three classification scenarios were evaluated, reflecting a progressive increase in feature richness:

\begin{itemize}
    \item \textbf{Solo only:} DTW consistency per joint group from the solo imitation condition.
    \item \textbf{Duo only:} DTW consistency per joint group from the duo imitation condition.
    \item \textbf{Solo + Duo:} DTW consistency from both  conditions combined with SCSI per joint group.
\end{itemize}

Leave-One-Subject-Out (LOSO) cross-validation was employed, treating each participant as an independent group and yielding one prediction per participant. This provides an unbiased estimate of generalization performance given the limited sample size. Hyperparameters ($C$ and $\gamma$) were tuned exclusively within each training fold using nested 3-fold stratified cross-validation over a grid of $C \in \{0.01, 0.1,
1, 10, 100\}$ and $\gamma \in \{\text{scale}, \text{auto}, 0.001, 0.01, 0.1\}$. Models were optimized to maximize balanced accuracy during the grid search. Classification performance is reported as balanced accuracy, sensitivity, and specificity.

\section{RESULTS}
\label{sec:results}

\subsection{Movement Onset Timing}
\label{sec:rt}
 
Reaction time was computed across all three conditions to assess whether autistic and neurotypical adults differ in the speed of movement initiation following the auditory beep. The results are summarized in
Table~\ref{tab:reaction_time}.

\begin{table}[!tb]
\centering
\caption{Reaction Time by Condition (mean $\pm$ SD, milliseconds)}
\label{tab:reaction_time}
\renewcommand{\arraystretch}{1.2}
\begin{tabular}{lccc}
\toprule
\textbf{Condition} & \textbf{Clinical ($n$=12)} & \textbf{Control ($n$=18)} & \textbf{$p$} \\
\midrule
Body Shake & $303.08 \pm 142.59$ & $357.88 \pm 133.36$ & 0.319 \\
Solo     & $311.47 \pm 205.54$ & $320.68 \pm 129.91$ & 0.896 \\
Duo      & $377.69 \pm 247.76$ & $386.38 \pm 201.23$ & 0.923 \\
\bottomrule
\multicolumn{4}{l}{\footnotesize Welch's $t$-test; no significant differences} \\
\end{tabular}
\end{table}

As detailed in Table~\ref{tab:reaction_time}, movement onset timing did not significantly differ between the clinical and control groups across any of the experimental conditions. Although both groups showed a descriptive trend toward longer mean reaction times and greater variability in the duo condition (mean reaction times of 377.69\,ms and 386.38\,ms, with standard deviations exceeding 200\,ms), the primary finding is the substantial within-group variance observed in all conditions. This confirms that reaction time does not constitute a reliable group-level discriminator in this dataset.

Crucially, this high variability in onset timing directly informed the parameters of the subsequent DTW consistency analysis. Including the initial frames of the execution window would conflate true kinematic variability with individual differences in movement initiation delays. Therefore, to ensure that the DTW consistency measures reflect ongoing movement rather than onset timing artifacts, the DTW distances were strictly computed over the final three seconds of the execution period, after all participants had fully transitioned into sustained movement.

\subsection{Global Movement Consistency}
\label{subsec:globaldtw}

Movement consistency across trials was quantified using DTW, computed over the execution period. Lower DTW distance indicates more consistent repetition of the movement pattern across trials.
Table~\ref{tab:dtw_global} summarizes the group comparison results across the three conditions. 

\begin{table}[!b]
\centering
\caption{Global DTW Distance by Condition (mean $\pm$ SD)}
\label{tab:dtw_global}
\renewcommand{\arraystretch}{1.2}
\begin{tabular}{lcccc}
\toprule
\textbf{Condition} & \textbf{Clinical} & \textbf{Control} & \textbf{$p$} & \textbf{$d$} \\
                   & \textbf{($n$=12)} & \textbf{($n$=18)} &              &              \\
\midrule
Body Shake & $149.8 \pm 33.9$ & $160.7 \pm 53.1$ & 0.485 & $-0.25$ \\
Solo     & $244.3 \pm 81.4$ & $265.9 \pm 140.7$ & 0.816 & $-0.19$ \\
Duo      & $284.1 \pm 80.2$ & $399.9 \pm 174.4$ & 0.066 & $-0.85$ \\
\bottomrule
\multicolumn{5}{l}{\footnotesize Mann-Whitney U test; $d$ = Cohen's $d$}
\end{tabular}
\end{table}

No significant difference was observed during the baseline body shake condition, suggesting that the two groups are broadly comparable in general motor behavior when no imitation is required. Similarly, no significant difference emerged during solo dance imitation.
A markedly different pattern was observed during duo dance imitation. Although the Mann-Whitney test did not reach conventional significance, the effect size was large ($d=-0.85$), with neurotypical controls showing substantially higher DTW distances (mean$=399.9$, SD$=174.4$) compared to autistic adults (mean$=284.1$, SD$=80.2$). 
This pattern indicates a trend toward increased movement variability in neurotypical individuals in response to the socially-framed stimulus, while autistic adults tend to maintain a consistent movement pattern regardless of social context, motivating the SCSI introduced in Section~\ref{sec:scsi}.

\subsection{Movement Consistency by Joint Group and Condition}

To determine specific anatomical regions driving group differences in movement consistency, DTW distances were analyzed across three distinct joint groups: upper body (shoulders, arms, hands), core (hip, abdomen, chest, neck, head), and lower body (thighs, shins, feet). This analysis was conducted independently for each imitation condition, with results summarized in Table~\ref{tab:dtw_jointgroup}.

\begin{table}[!b]
\centering
\caption{DTW Consistency by Joint Group and Condition (mean $\pm$ SD)}
\label{tab:dtw_jointgroup}
\renewcommand{\arraystretch}{1.2}

\begin{tabular}{llcccc}
\toprule
\textbf{Group} & \textbf{Cond.} & \textbf{Clinical} & \textbf{Control} & \textbf{$p$} & \textbf{$d$} \\
\midrule
\multirow{2}{*}{Upper} & Solo & $161.8 \pm 66.1 $ & $176.4 \pm 94.6$ & 0.719 & $-0.173$ \\
                       & Duo  & $187.9 \pm 57.3$ & $266.9 \pm 108.8$ & \textbf{0.044} & $-0.879$ \\
\addlinespace
\multirow{2}{*}{Core}  & Solo & $92.3 \pm 32.0$  & $103.3 \pm 66.7$  & 0.882 & $-0.205$ \\
                       & Duo  & $115.7 \pm 38.8$ & $153.7 \pm 79.8$ & 0.363 & $-0.588$ \\
\addlinespace
\multirow{2}{*}{Lower} & Solo & $147.5 \pm 48.0$ & $155.6 \pm 88.0$ & 0.751 & $-0.111$ \\
                       & Duo  & $167.6 \pm 47.9$ & $242.7 \pm 117.3$ & 0.066 & $-0.812$ \\
\bottomrule
\multicolumn{6}{l}{\footnotesize Mann-Whitney U test; $d$ = Cohen's $d$; \textbf{bold} = $p < 0.05$}
\end{tabular}
\end{table}

Consistent with the global DTW analysis (Section~\ref{subsec:globaldtw}), no significant group differences were detected during solo dance imitation across any joint group (all $p > 0.70$, $|d| < 0.21$). However, during the socially-framed duo dance imitation, a significant divergence emerged in the upper body ($p = 0.044$, $d = -0.879$). Neurotypical controls exhibited substantially higher DTW distances than autistic adults, reflecting increased movement variability in response to the social framing. A similar, but marginal, trend was observed in the lower body ($p = 0.066$, $d = -0.812$). Core body movements remained comparatively stable, showing no significant group differences in either condition (all $p > 0.21$).

These findings suggest that the social context of an imitation task selectively modulates kinematic consistency in the extremities, while axial kinematic stability demonstrates relative invariance between groups.

\subsection{Social Context Sensitivity Index}
\label{sec:scsi}

\begin{figure}[!t]
\centering
\includegraphics[width=0.999\columnwidth]{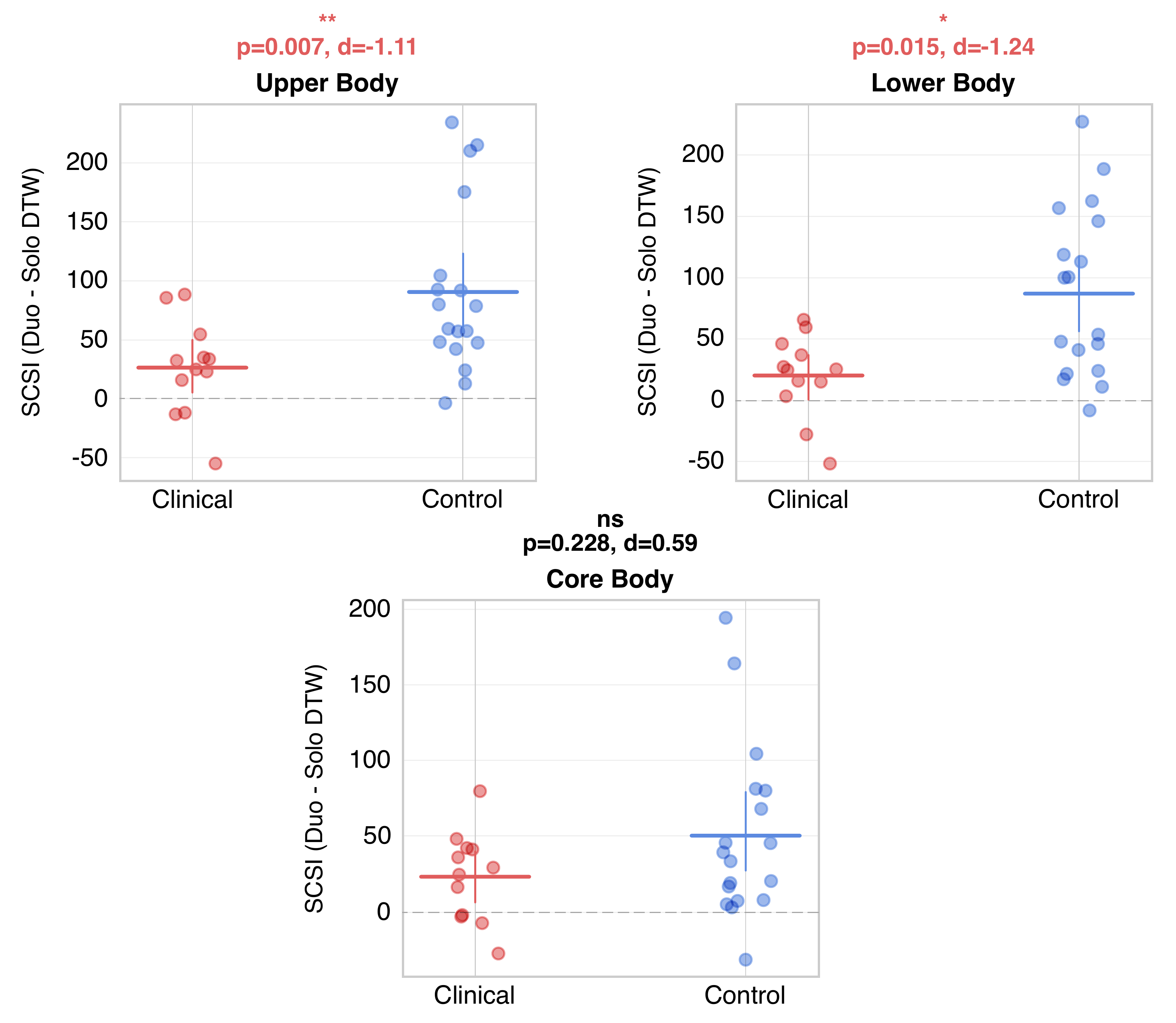}
\caption{Social Context Sensitivity Index (SCSI) per joint group. Each point represents one participant. Horizontal lines indicate group means; vertical bars indicate 95\% bootstrap confidence intervals. The dashed line at zero indicates no difference between conditions. Positive values indicate greater movement variability during duo than solo imitation. Statistical annotations: ** $p < 0.01$, * $p < 0.05$.}
\label{fig:scsi}
\end{figure}

The SCSI quantifies the within-subject shift in DTW consistency between the solo and duo conditions for each joint group (Eq.~\ref{eq:scsi}). Positive SCSI values reflect an increase in movement variability when transitioning to socially-framed imitation. Results are illustrated in Fig.~\ref{fig:scsi}.

Significant group differences in the SCSI were found in both the upper (Control: $90.5 \pm 70.1$ vs.\ Clinical: $26.2 \pm 38.9$; $p = 0.007$, $d = -1.107$) and lower extremities (Control: $87.0 \pm 65.2$ vs.\ Clinical: $20.1 \pm 32.2$; $p = 0.015$, $d = -1.238$). In these regions, neurotypical controls demonstrated markedly higher SCSI values, indicating a substantial increase in movement variability when adapting to the more complex social context. Conversely, autistic participants exhibited SCSI values closer to zero across all joint groups, demonstrating that their movement consistency was largely unaffected by social framing. Differences in core body SCSI (Control: $50.4 \pm 57.7$ vs.\ Clinical: $23.4 \pm 28.2$) did not reach statistical significance ($p = 0.228$, $d = -0.589$).

These large effect sizes ($|d| > 1.1$) highlight a robust behavioral divergence: the tendency to dynamically modulate motor variability in response to socially-framed biological motion is significantly attenuated in autistic adults, particularly in the limbs.

\subsection{Classification}
\label{sec:classification}

\begin{figure}[!t]
\centering
\includegraphics[width=\columnwidth]{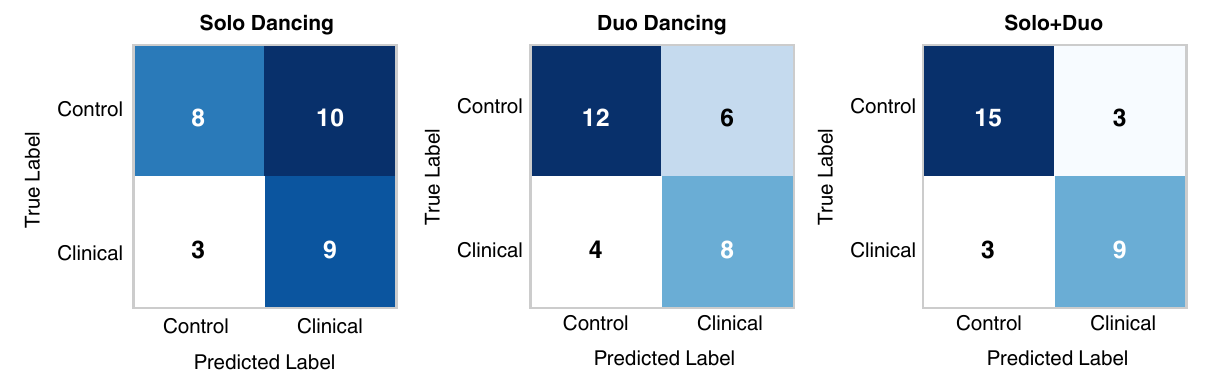}
\caption{Confusion matrix for each scenario evaluated. LOSO cross-validation, $n = 30$.}
\label{fig:confusion}
\end{figure}

Table~\ref{tab:classification} presents the classification performance across the three evaluated scenarios. 
Using only solo imitation features yielded a balanced accuracy of 59.7\% (sensitivity = 75.0\%, specificity = 44.4\%), highlighting that motor kinematics alone, without interpersonal social framing, provide limited discriminative power. Duo imitation features improved accuracy to 66.7\% (sensitivity = 66.7\%, specificity = 66.7\%), mirroring our statistical results that group differences manifest primarily during socially-framed tasks.
Optimal performance was achieved using the combined feature set (DTW consistencies from both conditions coupled with the SCSI). This model reached a balanced accuracy of 79.2\% (sensitivity = 75.0\%, specificity = 83.3\%), successfully identifying 9 of 12 autistic participants and 15 of 18 neurotypical participants (Fig.~\ref{fig:confusion}).

\begin{table}[!t]
\centering
\caption{Classification Performance by Scenario (LOSO, SVM)}
\label{tab:classification}
\renewcommand{\arraystretch}{1.2}
\begin{tabular}{lccc}
\toprule
\textbf{Condition} & \textbf{Bal. Acc.} & \textbf{Sensitivity} & \textbf{Specificity} \\
\midrule
Solo     & 59.7\% & 75.0\% & 44.4\% \\
Duo         & 66.7\% & 66.7\% & 66.7\% \\
Solo+Duo     & \textbf{79.2\%} & \textbf{75.0\%} & \textbf{83.3\%} \\
\bottomrule
\multicolumn{4}{l}{\footnotesize LOSO cross-validation; nested 3-fold CV for hyperparameter tuning.}
\end{tabular}
\end{table}

This progressive increase in classification accuracy, from solo (59.7\%) to duo (66.7\%) to the combined model (79.2\%), reinforces our primary hypothesis. Neither condition in isolation fully captures the group variance; rather, it is the dynamic modulation between solitary and social contexts (captured by the SCSI) that serves as the strongest discriminative signature.


\section{DISCUSSION}
\label{sec:discussion}

This study investigated whether motor imitation of dance sequences reveals differences in movement consistency between autistic and neurotypical adults, and whether social context modulates these differences. 

Analysis of movement onset revealed no significant differences between groups, characterized instead by high variability within the group. This aligns with the well-documented heterogeneity of motor profiles in autism, indicating that simple movement latency is not a reliable discriminative biomarker in this paradigm~\cite{Gowen2013}. 

The central finding of this study is that neurotypical controls showed significantly lower movement consistency during duo dance imitation compared to autistic adults, as captured by both the joint-group DTW analysis and the SCSI. Critically, this difference was absent during solo imitation and body shake, indicating that it is specific to the social context of the imitation stimulus rather than a general motor difference between groups.
While recent research has demonstrated that certain autistic subtypes exhibit reduced consistency during basic, non-social sensorimotor execution tasks~\cite{Mandelli2024}, our paradigm reveals a distinct kinematic signature specific to high-level scenarios. Because both groups in our study exhibited comparable movement consistency during the baseline and solo imitation conditions, the observed divergence in the duo condition is not a product of fundamental motor capability. Rather, the higher consistency maintained by the autistic group during duo imitation reflects an invariant, stereotyped execution of the motor plan in the presence of social stimuli. This sharply contrasts with the reduced movement consistency observed in neurotypical adults during duo imitation, reflected in higher DTW distances and positive SCSI values, indicating that their kinematics varied significantly more across trials when the stimulus carried social content. This variance reflects a natural tendency in neurotypical individuals to continuously modulate motor behavior in response to socially-framed biological motion, a process associated with social motor adaptation and perception-action coupling~\cite{Iacoboni2009, Cattaneo2007}. 
When interacting with a socially-embedded stimulus, neurotypical adults dynamically and flexibly adapt their motor execution, inherently introducing functional variability rather than rigidly replicating the movement across repetitions~\cite{Ordin2025, Wang2012
}. The effect was most pronounced in the upper and lower body, with no significant difference in the core body. This is consistent with the greater role of distal limbs in expressive and communicative movement, suggesting that social context modulation in neurotypical adults may be particularly evident in the body segments most associated with social gesture and expression.

Autistic adults, by contrast, maintained highly similar movement consistency across both conditions, yielding SCSI values close to zero across all joint groups. 
This reduced modulation of motor execution by social framing is consistent with previous accounts of differences in social orientation in autism \cite{Klin2003}.
Furthermore, it provides direct kinematic support for previous findings demonstrating that autistic individuals exhibit reduced top-down modulation of motor networks when processing and reacting to socially relevant biological motion~\cite{Nackaerts2012}.

The classification results directly reflected our statistical findings, reinforcing confidence in the discriminative value of social context contrast as a determinant in the motor signature of autism.
The achieved sensitivity of 75.0\%, specificity of 83.3\%, and accuracy of 79.2\% (Table IV) represent robust performance given the naturalistic setting, and the well-documented heterogeneity of motor presentations across the spectrum, and are consistent with prior work using computer vision or sensor-based motor assessments in autism, which typically report accuracies ranging from 73\% to 85\%~\cite{Vabalas2020,Zampella2021,Altozano2025,Ganai2025}. Furthermore, achieving high specificity is particularly valuable in this context, as it minimizes false positives, which is a critical requirement for any potential screening or assistive diagnostic tool to prevent over-diagnosis. 

Clinically, the identification of the SCSI holds promise as a quantitative biomarker, offering a potential objective tool for autism assessment that moves beyond subjective observational scales. Because this framework uses 3D motion capture and DTW, it holds potential for future translation into telerehabilitation platforms, eventually allowing clinicians to remotely monitor a patient’s motor-social adaptability over time. Regarding intervention, understanding that autistic individuals exhibit distinct, rigid motor execution during social imitation suggests a need to shift therapeutic goals. Future therapies could leverage our visual feedback framework to train functional variability, encouraging patients to explore adaptive movement strategies rather than focusing solely on rote motor repetition. Beyond these clinical applications, these findings also have significant implications for the design of human-centric intelligent systems, particularly in human-machine interaction (HMI) and socially assistive robotics. Because autistic individuals show distinct patterns of motor adaptation when interacting with socially-framed digital stimuli (in this case, point-light animations), future cybernetic systems and virtual agents must account for these neurodivergent profiles.

\section{CONCLUSION}
\label{sec:conclusion}

This work introduces a computational framework for characterizing motor imitation in autism using 3D motion capture with demonstrated potential for application in clinical and human-machine systems. By comparing kinematic consistency across solitary and socially-framed dance tasks, we observed a behavioral divergence: neurotypical adults tended to modulate their movement variability in response to social stimuli, whereas autistic adults showed stable movement consistency regardless of social context. This effect, quantified by the proposed SCSI, yielded large effect sizes in the upper ($d = -1.107$) and lower body ($d = -1.238$), and was absent during solo imitation and baseline conditions, suggesting it reflects a selective difference in social motor adaptation rather than general motor ability. 

Furthermore, our classification pipeline suggests that this social context contrast carries discriminative power, supporting the idea that the dynamic shift between non-social and social conditions is a stronger biomarker than performance in either isolated conditions.

In the broader context of human-centric intelligence, this work underscores the importance of integrating neurodivergent behavioral profiles into the development of interactive technologies. Future work will integrate these findings in larger samples to characterize the neural underpinnings of social context sensitivity in motor imitation, and investigate how these kinematic signatures can inform the design of inclusive human-machine interfaces, and potential medical systems.

\section*{Acknowledgments}
This work was supported by national funds through FCT - Foundation for Science and Technology, I.P., under the grants UID/00048/2025 (DOI: 10.54499/UIDB/00048/2025) and LA/P/0112/2020 (DOI: 10.54499/LA/P/0112/2020).

\bibliographystyle{IEEEtran}
\balance
\bibliography{ref.bib}

\begin{thebibliography}{10}
\providecommand{\url}[1]{#1}
\csname url@samestyle\endcsname
\providecommand{\newblock}{\relax}
\providecommand{\bibinfo}[2]{#2}
\providecommand{\BIBentrySTDinterwordspacing}{\spaceskip=0pt\relax}
\providecommand{\BIBentryALTinterwordstretchfactor}{4}
\providecommand{\BIBentryALTinterwordspacing}{\spaceskip=\fontdimen2\font plus
\BIBentryALTinterwordstretchfactor\fontdimen3\font minus \fontdimen4\font\relax}
\providecommand{\BIBforeignlanguage}[2]{{%
\expandafter\ifx\csname l@#1\endcsname\relax
\typeout{** WARNING: IEEEtran.bst: No hyphenation pattern has been}%
\typeout{** loaded for the language `#1'. Using the pattern for}%
\typeout{** the default language instead.}%
\else
\language=\csname l@#1\endcsname
\fi
#2}}
\providecommand{\BIBdecl}{\relax}
\BIBdecl

\bibitem{Lordan2021}
R.~Lordan, C.~Storni, and C.~A. de~Benedictis, ``Autism spectrum disorders: Diagnosis and treatment,'' \emph{Autism Spectrum Disorders}, pp. 17--32, 8 2021.

\bibitem{Vabalas2020}
A.~Vabalas, E.~Gowen, E.~Poliakoff, and A.~J. Casson, ``Applying machine learning to kinematic and eye movement features of a movement imitation task to predict autism diagnosis,'' \emph{Scientific Reports}, vol.~10, 5 2020.

\bibitem{Simeoli2024}
R.~Simeoli, A.~Rega, M.~Cerasuolo, R.~Nappo, and D.~Marocco, ``Using machine learning for motion analysis to early detect autism spectrum disorder: A systematic review,'' \emph{Review Journal of Autism and Developmental Disorders}, 3 2024.

\bibitem{Fournier2010}
K.~A. Fournier, C.~J. Hass, S.~K. Naik, N.~Lodha, and J.~H. Cauraugh, ``Motor coordination in autism spectrum disorders: A synthesis and meta-analysis,'' \emph{Journal of Autism and Developmental Disorders}, vol.~40, pp. 1227--1240, 10 2010.

\bibitem{Dowell2009}
L.~R. Dowell, E.~M. Mahone, and S.~H. Mostofsky, ``Associations of postural knowledge and basic motor skill with dyspraxia in autism: implication for abnormalities in distributed connectivity and motor learning,'' \emph{Neuropsychology}, vol.~23, pp. 563--570, 9 2009.

\bibitem{daSilva2025}
S.~H. da~Silva, M.~R. Felippin, L.~de~Oliveira~Medeiros, C.~Hedin-Pereira, and A.~A. Nogueira-Campos, ``A scoping review of the motor impairments in autism spectrum disorder,'' \emph{Neuroscience and Biobehavioral Reviews}, vol. 169, 1 2025.

\bibitem{Gowen2013}
E.~Gowen and A.~Hamilton, ``Motor abilities in autism: a review using a computational context,'' \emph{Journal of autism and developmental disorders}, vol.~43, pp. 323--344, 2 2013.

\bibitem{Duarte2022}
J.~V. Duarte, R.~Abreu, and M.~Castelo-Branco, ``A two-stage framework for neural processing of biological motion,'' \emph{NeuroImage}, vol. 259, 10 2022.

\bibitem{McEllin2018}
L.~McEllin, G.~Knoblich, and N.~Sebanz, ``Imitation from a joint action perspective,'' \emph{Mind and Language}, vol.~33, pp. 342--354, 9 2018.

\bibitem{Latrche2024}
K.~Latrèche, N.~Kojovic, I.~Pittet, S.~Natraj, M.~Franchini, I.~M. Smith, and M.~Schaer, ``Gesture imitation performance and visual exploration in young children with autism spectrum disorder,'' \emph{Journal of Autism and Developmental Disorders}, 2024.

\bibitem{Nackaerts2012}
E.~Nackaerts, J.~Wagemans, W.~Helsen, S.~P. Swinnen, and N.~Wenderoth, ``Recognizing biological motion and emotions from point-light displays in autism spectrum disorders,'' \emph{PLoS ONE}, vol.~7, p. 44473, 2012.

\bibitem{McCleery2013}
J.~P. McCleery, N.~A. Elliott, D.~S. Sampanis, and C.~A. Stefanidou, ``Motor development and motor resonance difficulties in autism: Relevance to early intervention for language and communication skills,'' \emph{Frontiers in Integrative Neuroscience}, vol.~7, 4 2013.

\bibitem{Yan2018}
\BIBentryALTinterwordspacing
S.~Yan, Y.~Xiong, and D.~Lin, ``Spatial temporal graph convolutional networks for skeleton-based action recognition,'' \emph{32nd AAAI Conference on Artificial Intelligence, AAAI 2018}, pp. 7444--7452, 1 2018. [Online]. Available: \url{http://arxiv.org/abs/1801.07455}
\BIBentrySTDinterwordspacing

\bibitem{Taye2023}
M.~M. Taye, ``Understanding of machine learning with deep learning: Architectures, workflow, applications and future directions,'' \emph{Computers}, vol.~12, 4 2023.

\bibitem{laraieee}
L.~Pereira, T.~Sousa, R.~Vigário, M.~Castelo-Branco, and J.~R. Paulo, ``Motion analysis in autism: Quantification and classification of dancing and walking tasks,'' in \emph{Proceedings of the 8th IEEE Portuguese Meeting on Bioengineering (ENBENG 2025)}.\hskip 1em plus 0.5em minus 0.4em\relax IEEE, 2025.

\bibitem{datasetarticle}
J.~R. Paulo, T.~Sousa, J.~Perdiz, L.~Pereira, M.~Vasen, S.~Mouga, G.~Pires, and M.~Castelo-Branco, ``A multimodal dataset addressing motor function in autism,'' \emph{Scientific Data}, vol.~12, 6 2025.

\bibitem{Liu2024}
Y.~Liu, Y.~A. Zhang, M.~Zeng, and J.~Zhao, ``A novel distance measure based on dynamic time warping to improve time series classification,'' \emph{Information Sciences}, vol. 656, p. 119921, 1 2024.

\bibitem{Barth2013}
J.~Barth, C.~Oberndorfer, P.~Kugler, D.~Schuldhaus, J.~Winkler, J.~Klucken, and B.~M. Eskofier, ``Subsequence dynamic time warping as a method for robust step segmentation using gyroscope signals of daily life activities,'' 2013.

\bibitem{Lee2024}
H.-S. Lee, J.-H. Lee, and K.-R. Kim, ``A method for selecting the optimal warping path of dynamic time warping in gait analysis,'' 2024.

\bibitem{Bolis2018}
D.~Bolis and L.~Schilbach, ``Observing and participating in social interactions: Action perception and action control across the autistic spectrum,'' \emph{Developmental Cognitive Neuroscience}, vol.~29, pp. 168--175, 1 2018.

\bibitem{Trujillo2018}
J.~P. Trujillo, I.~Simanova, H.~Bekkering, and A.~Özyürek, ``Communicative intent modulates production and comprehension of actions and gestures: A kinect study,'' \emph{Cognition}, vol. 180, pp. 38--51, 11 2018.

\bibitem{Iacoboni2009}
M.~Iacoboni, ``Imitation, empathy, and mirror neurons,'' \emph{Annual review of psychology}, vol.~60, pp. 653--670, 1 2009.

\bibitem{Cattaneo2007}
L.~Cattaneo, M.~Fabbri-Destro, S.~Boria, C.~Pieraccini, A.~Monti, G.~Cossu, and G.~Rizzolatti, ``Impairment of actions chains in autism and its possible role in intention understanding,'' \emph{Proceedings of the National Academy of Sciences of the United States of America}, vol. 104, pp. 17\,825--17\,830, 11 2007.

\bibitem{Forbes2016}
P.~A. Forbes, X.~Pan, and A.~F. Antonia, ``Reduced mimicry to virtual reality avatars in autism spectrum disorder,'' \emph{Journal of Autism and Developmental Disorders}, vol.~46, pp. 3788--3797, 12 2016.

\bibitem{Todorova2019}
G.~K. Todorova, R.~E. M.~B. Hatton, and F.~E. Pollick, ``Biological motion perception in autism spectrum disorder: A meta-analysis,'' \emph{Molecular Autism}, vol.~10, 12 2019.

\bibitem{co}
J.~Cook, ``From movement kinematics to social cognition: the case of autism.''

\bibitem{dstdoi}
Paulo, J. R., Sousa, T., Perdiz, J., Pereira, L., Vasen, M., Mouga, S., Pires, G., and Castelo-Branco, M. Move4AS: A Multimodal Dataset Addressing Motor Function in Autism. \textit{figshare} \url{https://doi.org/10.6084/m9.figshare.28296518}.

\bibitem{normalizationscale_Zanfir2013}
M.~Zanfir, M.~Leordeanu, and C.~Sminchisescu, ``The moving pose: An efficient 3d kinematics descriptor for low-latency action recognition and detection,'' in \emph{Proceedings of the IEEE International Conference on Computer Vision}.\hskip 1em plus 0.5em minus 0.4em\relax Institute of Electrical and Electronics Engineers Inc., 2013, pp. 2752--2759.

\bibitem{normalization_pos_orient_Sedmidubsky2017}
J.~Sedmidubsky, P.~Elias, and P.~Zezula, ``Effective and efficient similarity searching in motion capture data,'' \emph{Multimedia Tools and Applications}, vol.~77, pp. 12\,073--12\,094, 5 2017.

\bibitem{Brenner2019}
E.~Brenner and J.~B. Smeets, ``How can you best measure reaction times?'' \emph{Journal of Motor Behavior}, vol.~51, pp. 486--495, 9 2019.

\bibitem{Switonski2019}
A.~Switonski, H.~Josinski, and K.~Wojciechowski, ``Dynamic time warping in classification and selection of motion capture data,'' \emph{Multidimensional Systems and Signal Processing}, vol.~30, pp. 1437--1468, 7 2019.

\bibitem{Mandelli2024}
V.~Mandelli, I.~Landi, S.~B. Ceccarelli, M.~Molteni, M.~Nobile, A.~D’Ausilio, L.~Fadiga, A.~Crippa, and M.~V. Lombardo, ``Enhanced motor noise in an autism subtype with poor motor skills,'' \emph{Molecular Autism}, vol.~15, p.~36, 12 2024.

\bibitem{Ordin2025}
M.~Ordin, N.~Barbarroja, L.~Polyanskaya, H.~M. Manrique, and M.~Castelo-Branco, ``Metacognition and cognitive flexibility in autistic and neurotypically-developing populations,'' \emph{Brain and behavior}, vol.~15, 7 2025.

\bibitem{Wang2012}
Y.~Wang and A.~F. de~Hamilton, ``Social top-down response modulation (storm): a model of the control of mimicry in social interaction,'' \emph{Frontiers in Human Neuroscience}, vol.~6, p. 153, 6 2012.

\bibitem{Klin2003}
A.~Klin, W.~Jones, R.~Schultz, and F.~Volkmar, ``The enactive mind, or from actions to cognition: lessons from autism,'' \emph{Philosophical transactions of the Royal Society of London. Series B, Biological sciences}, vol. 358, pp. 345--360, 2 2003.

\bibitem{Zampella2021}
C.~J. Zampella, E.~Sariyanidi, A.~G. Hutchinson, G.~K. Bartley, R.~T. Schultz, and B.~Tunç, ``Computational measurement of motor imitation and imitative learning differences in autism spectrum disorder,'' in \emph{ICMI 2021 Companion - Companion Publication of the 2021 International Conference on Multimodal Interaction}.\hskip 1em plus 0.5em minus 0.4em\relax Association for Computing Machinery, Inc, 10 2021, pp. 362--370.

\bibitem{Altozano2025}
A.~Altozano, M.~E. Minissi, M.~Alcañiz, and J.~Marín-Morales, ``Introducing 3dcnn resnets for asd full-body kinematic assessment: A comparison with hand-crafted features,'' \emph{Expert Systems with Applications}, vol. 270, 1 2025.

\bibitem{Ganai2025}
U.~J. Ganai, A.~Ratne, B.~Bhushan, and K.~S. Venkatesh, ``Early detection of autism spectrum disorder: gait deviations and machine learning,'' \emph{Scientific Reports}, vol.~15, 1 2025.

\end{thebibliography}
\balance

\end{document}